\documentclass{article}

\usepackage{amsfonts}

\usepackage[dvips]{graphicx}
\usepackage{algorithm}
\usepackage{algorithmic}
\usepackage{amssymb}
\usepackage{color}
\usepackage{fancyhdr}
\usepackage{natbib}
\usepackage{amsmath} 
\usepackage{amssymb}  
\usepackage{mathrsfs} 
\usepackage{stmaryrd}

\begin{document}
\newcommand{\fhat}{\hat{f}}
\newcommand{\Fhat}{\hat{F}}
\newcommand{\Fhatbar}{ {\hat{\underline{F}}} }
\newcommand{\Fbar}{ {\underline{F}} }
\newcommand{\var}{\mathbb{V}}
\newcommand{\mean}{\mathbb{E}}
\newcommand{\xith}{\x^{(i)}}

\bibliographystyle{unsrtnat}

\title{Bias-Variance Trade-offs: Novel Applications}
\author{Dev Rajnarayan \and David Wolpert} 
\maketitle

\section*{Synonyms}
Bias-variance trade-offs, bias plus variance.

\section*{Definition}
Consider a given random variable $\Fbar$ and a random variable
that we can modify, $\Fhatbar$. We wish to use a sample of $\Fhatbar$
as an estimate of a sample of $\Fbar$. The mean squared error
between such a pair of samples is a sum of four terms. The first term
reflects the statistical coupling between $\Fbar$ and $\Fhatbar$ and
is conventionally ignored in bias-variance analysis. The second term
reflects the inherent noise in $\Fbar$ and is independent of the
estimator $\Fhatbar$. Accordingly, we cannot affect this term. In
contrast, the third and fourth terms depend on $\Fhatbar$. The third
term, called the bias, is independent of the precise samples of both
$\Fbar$ and $\Fhatbar$, and reflects the difference between the means
of $\Fbar$ and $\Fhatbar$. The fourth term, called the variance, is
independent of the precise sample of $\Fbar$, and reflects the
inherent noise in the estimator as one samples it.  These last two
terms can be modified by changing the choice of the estimator. In
particular, on small sample sets, we can often decrease our mean
squared error by, for instance, introducing a small bias that causes a
large reduction the variance. While most commonly used in machine
learning, this article shows that such bias-variance trade-offs are
applicable in a much broader context and in a variety of
situations. We also show, using experiments, how existing bias-variance trade-offs can be applied in novel circumstances to improve the performance of a class of optimization algorithms.

\section*{Motivation and Background}
\label{sec:BPVDerivation}
In its simplest form, the bias-variance decomposition is based on the
following idea. Say we have a Euclidean random variable $\Fbar$ taking
on values $F$ distributed according to a density function $p(F)$.  We
want to estimate what value we would get if were to sample
$p(F)$. However we do not (or cannot) do this simply by sampling
$\Fbar$ directly. Instead, to form our estimate, we sample a different
Euclidean random variable $\Fhatbar$ taking on values $\Fhat$
distributed according to $p(\Fhat)$. Assuming a quadratic loss
function, the quality of our estimate is measured by its Mean Squared
Error (MSE):
\begin{eqnarray}
\textrm{MSE}(\Fhatbar) &\equiv& \int p(\Fhat, F) \,(\Fhat - F)^2 d\Fhat dF.
\label{eq:bpv}
\end{eqnarray} 

$ $

\noindent {\bf{Example 1:}} To illustrate Eq.~\ref{eq:bpv}, consider the
simplest type of supervised machine learning problem, where there is a
finite input space $X$, the output space $Y$ is real numbers, and
there is no noise. In such learning there is some deterministic
`target function' $f$ that maps each element of $X$ to a single
element of $Y$.  There is a `prior' probability density function
$\pi(f)$ over target functions, and it gets sampled to produce some
particular target function, $f$.  Next, $f$ is IID sampled at a set of
$m$ inputs to produce a `training set' ${\cal{D}}$ of input-output
pairs.

\noindent For simplicity, say there is some single fixed ``prediction point'' $x
\in X$.  Our  goal in supervised learning is to estimate
$f(x)$. However $f$ is not known to us. Accordingly, to perform the
estimation the training set is presented to a `learning algorithm',
which in response to the training set produces a guess $g(x)$ for the
value $f(x)$.

\noindent This entire stochastic procedure defines a joint distribution $\pi(f,
{\cal{D}}, f(x), g(x))$. We can marginalize it to get a distribution
$\pi(f(x), g(x))$. Since $g(x)$ is supposed to be an estimate of
$f(x)$, we can identify $g(x)$ as the value $\Fhat$ of the random
variable $\Fhatbar$ and $f(x)$ as the value $F$ of $\Fbar$. In other
words, we can define $p(F, \Fhat) = \pi(f(x), g(x))$.
If we now ask what the mean squared error is of the guess made by our
learning algorithm for the value $f(x)$, we get Eq.~\ref{eq:bpv}.

\noindent Note that one would expect that this $\Fbar$ and $\Fhatbar$ are
statistically dependent (Indeed, if they weren't dependent, then the
dependence of thelearning algorithm on $\cal{D}$ would be pointless.)
Formally, the dependence can be established by writing
\begin{eqnarray*}
p(f(x), g(x)) &=& \int d{\cal{D}} \; p(f(x), g(x) \mid {\cal{D}}) \;
p({\cal{D}}) \\
&=& \int d{\cal{D}} \; p(g(x) \mid f(x), {\cal{D}}) \;p(f(x) \mid
{\cal{D}}) \; p({\cal{D}}) \\
&=& \int d{\cal{D}} \; p(g(x) \mid {\cal{D}}) \; p(f(x) \mid
{\cal{D}}) \; p({\cal{D}})
\end{eqnarray*}
(since the guess of the learning algorithm is determined in full by
the training set), and then noting that in general this integral
differs from the product
\begin{eqnarray*}
p(f(x)) \; p(g(x)) &=& \left[\int d{\cal{D}} \; p(f(x) \mid {\cal{D}})
p({\cal{D}})\right]\left[\int d{\cal{D}} \; p(g(x) \mid {\cal{D}})
p({\cal{D}})\right].
\end{eqnarray*}

$ $

In Ex. 1 $\Fbar$ and $\Fhatbar$ are statistically coupled. Such
coupling is extremely common. In practice though, such coupling is
simply ignored in analyses of bias plus variance, without any
justification. In particular Bayesian supervised learning avoids any
explicit consideration of bias plus variance. For its part,
non-Bayesian supervised learning avoids consideration of the coupling
by replacing the distribution $p(F,
\Fhat)$ with the associated product of marginals, $p(F)p(\Fhat)$.  For
now we follow that latter practice. So our equation for MSE reduces to
\begin{eqnarray}
\textrm{MSE}(\Fhatbar) &=& \int p(\Fhat) p(F) \,(\Fhat - F)^2 d\Fhat dF.
\label{eq:exp_cost}
\end{eqnarray} 
(If we were to account for the coupling of $\Fhat$ and $\Fhatbar$ an
additive correction term would need to be added to the right-hand side. For instance, see
\citet{wolp97}.)

Using simple algebra, the right hand side of Eq.~\ref{eq:exp_cost} can
be written as the sum of three terms. The first is the variance of
$\Fbar$. Since this is beyond our control in designing the
estimator $\Fhatbar$, we ignore it for the rest of this
article. The second term involves a mean that describes the
deterministic component of the error. This term depends on both the
distribution of $\Fbar$ and that of $\Fhatbar$, and quantifies how close the
means of those distributions are.  The third term is a
variance that describes stochastic variations from one sample to the
next. This term is independent of the random variable being
estimated. Formally, up to an overall additive constant, we can write
\begin{eqnarray}
\textrm{MSE}(\Fhat) &=& \int p(\Fhat) (\Fhat^2 - 2 F \Fhat + F^2) d\Fhat,\nonumber\\
&=& \int p(\Fhat) \Fhat^2 d\Fhat - 2 F \int p(\Fhat) \Fhat\, d\Fhat + F^2,\nonumber\\ 
&=& \overbrace{\var(\Fhat) + [\mean(\Fhat)]^2} - 2 F \;\mean(\Fhat) + F^2,\nonumber\\
&=& \var({\Fhat}) + \underbrace{[F - \mean(\Fhat)]^2}, \nonumber\\
&=& \textrm{variance} + \textrm{bias}^2.
\label{eq:biasPlusVariance}
\end{eqnarray}

In light of Eq.~\ref{eq:biasPlusVariance}, one way to try to reduce
afexpected quadratic error is to modify an estimator to trade-off bias
and variance. Some of the most famous applications of such
bias-variance trade-offs occur in parametric machine learning, where
many techniques have been developed to exploit the trade-off. However
there are some extensions of that trade-off that could be applied in
parametric machine learning that have been ignored by the
community. We illustrate one of them here.

Moreover, the bias-variance trade-off arises in many other fields
besides parameteric machine learning. In particular, as we illustrate
here, it arises in integral estimation and optimization. In the rest
of this paper we present some novel applications of the bias-variance
trade-off, and describe some interesting features in each case. A
recurring theme is that whenever a bias-variance trade-off arises in a
particular field, we can use many techniques from parametric machine
learning that have been developed for exploiting this trade-off. The
novel applications of the tradeoff discussed here are instances of the
Probability Collectives (PC) \cite{wora07,wost06,wobi04a,
wobi04b,mawo05}, a general approach to using probability distributions
to do blackbox optimization.

\section*{Applications}

In this section, we describe some applications of the bias-variance
tradeoff. First, we describe Monte Carlo (MC) techniques for the
estimation of integrals, and provide a brief analysis of bias-variance
trade-offs in this context. Next, we introduce the field of Monte
Carlo Optimization (MCO), and illustrate that there are more
subtleties involved than in simple MC. Then, we describe the field of
Parametric Machine Learning, which, as will show, is formally
identical to MCO. Finally, we present an application of Parametric
Learning (PL) techniques to improve the performance of MCO
algorithms. We do this in the context of an MCO problem that is
central to how PC addresses black-box optimization.

\subsection*{Monte Carlo Estimation of Integrals Using Importance Sampling}

Monte Carlo methods are often the method of choice for estimating
difficult high-dimensional integrals. Consider a function $f \colon
X \rightarrow \mathbb{R}$, which we want to integrate over
some region $\mathcal{X} \subseteq X$, yielding the value
$F$, as given by
\begin{equation*}
F = \int_{\mathcal{X}} dx\, f(x).
\end{equation*} 
We can view this as a random variable $\Fbar$, with density function given
by a Dirac delta function centered on $F$.  Therefore, the variance of $\Fbar$
is 0, and Eq.~\ref{eq:biasPlusVariance} is exact.

A popular MC method to estimate this integral is importance sampling
\cite[see][]{roca04}. This exploits the law of large numbers as follows:
i.i.d. samples $x^{(i)}, \; i=1,\ldots, m$ are generated from a
so-called importance distribution $h(x)$ that we control, and the
associated values of the integrand, $f(x^{(i)})$ are computed. Denote
these `data' by 
\begin{equation}
\mathcal{D} = \{(x^{(i)}, f(x^{(i)}), \;i=1,\ldots,m\}.
\label{eq:data}
\end{equation} 
Now,
\begin{eqnarray*}
F &=& \int_{\mathcal{X}} dx\, h(x) \frac{f(x)}{h(x)},\nonumber\\
&=& \lim_{m \rightarrow \infty} \frac{1}{m}\sum_{i=1}^m \frac{f(x^{(i)})}{h(x^{(i)})}\textrm{ with probability 1.}
\end{eqnarray*}

Denote by $\Fhatbar$ the random variable with value given by the
sample average for $\mathcal{D}$:
\begin{equation*}
\Fhat = \frac{1}{m}\sum_{i=1}^m \frac{f(x^{(i)})}{h(x^{(i)})}.
\end{equation*}
We use $\Fhatbar$ as our statistical estimator for $\Fbar$, as we
broadly described in the introductory section.  Assuming a quadratic
loss function, $L(\hat{F}, F) = (F - \hat{F})^2$, the bias-variance decomposition described
in Eq.~\ref{eq:biasPlusVariance} applies \emph{exactly}. 
It can be shown that the estimator $\Fhatbar$ is unbiased, that is,
$\mean(\Fhatbar) = F$, where the mean is over samples of $h$. Consequently, the MSE of this estimator is just its variance. The
choice of sampling distribution $h$ that minimizes this variance is
given by~\citep*[see][]{roca04}
\begin{equation*}
h^{\star}(x) = \frac{|f(x)|}{\int_{\mathcal{X}} |f(x')| dx'}.
\end{equation*}

By itself, this result is not very helpful, since the equation for the
optimal importance distribution contains a similar integral to the one
we are trying to estimate. For non-negative integrands $f(x)$, the
VEGAS algorithm \citep{lepa78} describes an adaptive method to find
successively better importance distributions, by iteratively
estimating $\Fbar$, and then using that estimate to generate the next
importance distribution $h$. In the case of these unbiased estimators,
there is no trade-off between bias and variance, and minimizing MSE is
achieved by minimizing variance.

\subsection*{Monte Carlo Optimization}
Instead of a \emph{fixed} integral to evaluate, consider a parametrized integral
\begin{equation*}
F(\theta) = \int_{\mathcal{X}} dx\,f_{\theta}(x).
\end{equation*}
Further, suppose we are interested in finding the value of the
parameter $\theta \in \Theta$ that minimizes $F(\theta)$: 
\begin{equation*}
\theta^{\star} = \arg\min_{\theta \in \Theta} F(\theta).
\end{equation*}
In the case where the functional form of $f_{\theta}$ is not explicitly known,
one approach to solve this problem is a technique called Monte Carlo
Optimization (MCO) \citep*[see][]{erno98}, involving repeated MC estimation of the integral in question with adaptive
modification of the parameter $\theta$. 

We proceed by analogy to the case with MC. First, we introduce the
$\theta$-indexed random variable $\Fbar(\theta)$, all of whose
components have delta-function distributions about the associated
values $F(\theta)$. Next, we introduce a $\theta$-indexed vector
random variable $\Fhatbar$ with values 
\begin{equation}
{\Fhat} \equiv \{{\Fhat}(\theta)\; \forall\,\theta \in \Theta\}.
\end{equation}
Each real-valued component ${\Fhatbar}(\theta)$ can be sampled and
viewed as an estimate of $\Fbar(\theta)$. 


For example, let $\mathcal{D}$ be a data set as described in Eq.~\ref{eq:data}. Then for every $\theta$, any
sample of $\mathcal{D}$ provides an associated estimate 
\begin{equation*}
\hat{F}(\theta) = \frac{1}{m}\sum_{i=1}^m \frac{f_\theta(x^{(i)})}{h(x^{(i)})}.
\end{equation*}  
That average serves as an estimate of $\Fbar(\theta)$. Formally, $\Fhatbar$ is a function of the
random variable $\mathcal{D}$, and is given by such averaging over the
elements of $\mathcal{D}$. So, a sample of $\mathcal{D}$ provides a
sample of $\Fhatbar$. {\it{A priori}}, we make no restrictions on $\Fhatbar$, and so, in general, its components may be
statistically coupled with one another. Note that this coupling arises even though we are, for simplicity, treating each function $\Fbar(\theta)$ as having a
delta-function distribution, rather than as having a non-zero variance that would reflect
our lack of knowledge of the $f(\theta)$ functions.


However $\Fhatbar$ is defined, given a sample of $\Fhat$, one way
to estimate $\theta^{\star}$ is
\begin{equation*}
\hat{ {{\theta}}} ^{\star} = \arg\min_{\theta \in \Theta} \hat{F}(\theta)
\end{equation*}
We call this approach `natural' MCO. As an example, say that
$\mathcal{D}$ is a set of $m$ samples of $h$, and let
\begin{equation*}
\Fhat(\theta) \triangleq
\frac{1}{m}\sum_{i=1}^m \frac{f_{\theta}(x^{(i)})}{h(x^{(i)})},
\end{equation*} 
as above. Under this choice for $\Fhatbar$,
\begin{equation}
\hat{\theta}^{\star} =\arg\min_{\theta \in \Theta}\frac{1}{m}\sum_{i=1}^m \frac{f_{\theta}(x^{(i)})}{h(x^{(i)})}.
\label{eq:naiveMCO}
\end{equation}
We call this approach `naive' MCO.

Consider \emph{any} algorithm that estimates $\theta^{\star}$ as a
single-valued function of ${\Fhat}$. The estimate of $\theta^\star$ produced by that
algorithm is itself a random variable, since it is a function of the
random variable $\Fhatbar$. Call this random variable $\hat{{\underline{\theta}}}^{\star}$, taking on
values $\hat{ {\theta}}^{\star}$. Any MCO algorithm is defined by
$\hat{ {\underline{\theta}}}^{\star}$; that random variable
encapsulates the output estimate made by the algorithm.

To analyze the error of such an algorithm, consider the associated
random variable given by the true parametrized integral $F(\hat{{\underline{\theta}}}^{\star})$. The
difference between a sample of $F({\hat{{\underline{\theta}}}}^{\star})$ and the
true minimal value of the integral, $F({\theta^{\star}}) =
\min_\theta F(\theta)$, is the error introduced by our
estimating that optimal $\theta$ as a sample of
${\hat{{\underline{\theta}}}}^{\star}$.  Since our aim in MCO is to 
minimize $F(\theta)$, we adopt the loss function $L({\hat{{\underline{\theta}}}}^{\star}, {\theta^{\star}})
\triangleq F({\hat{{\underline{\theta}}}}^{\star}) - F({\theta^{\star}})
$. This is in contrast to our discussion on MC integration, which involved quadratic loss. The current loss function just equals
$F({\hat{{\underline{\theta}}}}^{\star})$ up to an additive constant
$F({\theta^{\star}})$ that is fixed by the MCO problem at
hand and is beyond our control. Up to that additive constant, the associated expected loss is
\begin{eqnarray}
\mean(L) &=& \int d\hat{\theta}^{\star} p(\hat{\theta}^{\star}) F(\hat{\theta}^{\star}).
\label{eq:expectedLoss}
\end{eqnarray}
Now change coordinates in this integral from the values of the scalar
random variable $\hat{ {\underline{\theta}}}^{\star}$ to the values of the underlying
vector random variable ${\Fhatbar}$.  The expected loss now becomes
\begin{eqnarray*}
\mean(L) &=& \int d{\Fhat} \;p({\Fhat})
F({\hat{\theta}}^\star(\Fhat)).
\end{eqnarray*}

The natural MCO algorithm provides some insight into these
results.  For that algorithm,
\begin{eqnarray}
\mean(L) &=& \int d{\Fhat} \;p({\Fhat})
F(\arg\min_{\theta}{\Fhat}(\theta)) \nonumber \\
&=& \int d{\Fhat}(\theta_1) d\Fhat(\theta_2) \ldots \;p({\Fhat(\theta_1),
\Fhat(\theta_2), \ldots}) F(\arg\min_{\theta}{\Fhat}(\theta)).
\label{eq:natural_loss}
\end{eqnarray}
For any fixed $\theta$, there is an error between samples of
$\Fhat(\theta)$ and the true value $F(\theta)$. Bias-variance
considerations apply to this error, exacty as in the discussion of MC
above. We are not, however, concerned with $\Fhat$ for a single
component $\theta$, but rather for a set $\Theta$ of $\theta$'s. 

The simplest such case is where the components of $\Fhat(\Theta)$ are independent. Even so, $\arg\min_\theta \Fhat(\theta)$ is distributed according
to the laws for extrema of multiple independent random
variables, and this distribution depends on higher-order moments of each random variable
$\Fhat(\theta)$. This means that $\mean[L]$ also depends on such higher-order
moments. Only the first two moments, however, arise in the bias and
variance for any single $\theta$. Thus, even in the simplest possible case, the bias-variance considerations for the individual $\theta$ do not provide a complete analysis.

In most cases, the components of $\Fhat$ are $not$ independent. Therefore, in order to analyze $\mean[L]$, in addition to higher moments of the distribution for each $\theta$, we must now also consider higher-order moments coupling the estimates
${\Fhat}(\theta)$ for different $\theta$. 

Due to these effects, it may be quite acceptable for all the
components ${\Fhat}(\theta)$ to have both a large bias and a large
variance, as long as they still order the $\theta$'s correctly with
respect to the true $F(\theta)$. In such a situation, large
covariances could ensure that if some ${\Fhat}(\theta)$ were incorrectly large, then $\Fhat(\theta'), \theta' \neq \theta$ would also be incorrectly large. This coupling between the components of $\Fhat$ would preserve the ordering of $\theta$'s under $\Fbar$. So, even with large bias and variance for each $\theta$, the estimator as a whole would still work well.

Nevertheless, it $is$ sufficient to design estimators
${\Fhat}(\theta)$ with sufficiently small bias plus variance for
each single $\theta$. More precisely, suppose that those terms are
very small on the scale of differences $F(\theta) - F(\theta')$ for
any $\theta$ and $\theta'$. Then by Chebychev's inequality, we know that
the density functions of the random variables $\Fhatbar(\theta)$ and
$\Fhatbar(\theta')$ have almost no overlap.  Accordingly, the probability that a sample of $\Fhatbar(\theta) - \Fhatbar(\theta')$ has the opposite sign of $F(\theta) -
F(\theta')$ is almost zero.

Evidently, $\mean[L]$ is generally determined by a complicated
relationship involving bias, variance, covariance, and higher
moments. Natural MCO in general, and naive MCO in particular, ignore all of these effects, and consequently, 
often perform quite poorly in practice. In the next section we discuss some ways of
addressing this problem.

\subsection*{Parametric Machine Learning}

There are many versions of the basic MCO problem described in the
previous section. Some of the best-explored arise in parametric
density estimation and parametric supervised learning, which together
comprise the field of Parametric machine Learning (PL).

In particular, parametric supervised learning attempts to solve
\begin{equation*}
\arg\min_{\theta \in \Theta} \int dx\, p(x) \int dy \,p(y\mid
x)f_{\theta}(x).
\end{equation*}
Here, the values $x$ represent inputs, and the values $y$ represent
corresponding outputs, generated according to some stochastic process
defined by a set of conditional distributions $\{p(y\mid x),\;
x\in\mathcal{X}\}$. Typically, one tries to solve this problem
by casting it as an MCO problem, For instance, say we adopt a
quadratic loss between a predictor $z_{\theta}(x)$ and the true value
of $y$. Using MCO notation, we can express the associated
supervised learning problem as finding $\arg\min_\theta F(\theta)$,
where
\begin{eqnarray}
l_{\theta}(x) &=& \int dy\, p(y\mid x)\, (z_{\theta}(x) - y)
^2,\nonumber\\
f_{\theta}(x) &=& p(x)\, l_{\theta}(x),\nonumber\\
F(\theta) &=& \int dx\, f_{\theta}(x).
\label{eq:sup_learn_1}
\end{eqnarray}

Next, the argmin is estimated by minimizing a sample-based estimate of
the $F(\theta)$'s. More precisely, we are given a `training set' of
samples of $p(y \mid x)\,p(x)$, \{$(x^{(i)}, y^{i}) i = 1, \ldots,
m$\}. This training set provides a set of associated estimates of
$F(\theta)$:
\begin{equation*}
\hat{F}(\theta) = \frac{1}{m}\sum_{i=1}^m l_{\theta}(x^{(i)}).
\end{equation*} 
These are used to estimate $\arg\min_\theta F(\theta)$, exactly as in
MCO.  In particular, one could estimate the minimizer of $F(\theta)$
by finding the minimium of $\hat{F}(\theta)$, just as in natural
MCO. As mentioned above, this MCO algorithm can perform very poorly in
practice. In PL, this poor performance is called `overfitting the
data'.

There are several formal approaches that have been explored in PL to
try to address this `overfitting the data'. Interestingly, none
are based on direct consideration of the random variable
$F({\hat{\theta}}^\star(\Fhatbar))$ and the ramifications of its
distribution for expected loss (cf.  Eq.~\ref{eq:natural_loss}). In
particular, no work has applied the mathematics of extrema of multiple
random variables to analyze the bias-variance-covariance trade-offs
encapsulated in Eq.~\ref{eq:natural_loss}.

The PL approach that perhaps comes closest to such direct
consideration of the distribution of
$F({\underline{\hat{\theta}}}^\star)$ is uniform convergence theory,
which is a central part of Computational Learning Theory
\citep[see][]{angl92}. Uniform convergence theory starts by crudely
encapsulating the quadratic loss formula for expected loss under
natural MCO, Eq.~\ref{eq:natural_loss}. It does this by considering
the worst-case bound, over possible $p(x)$ and $p(y \mid x)$, of the
probability that $F({\underline{\theta}}^\star)$ exceeds $\min_\theta
F(\theta)$ by more than $\kappa$. It then examines how that bound
varies with $\kappa$. In particular, it relates such variation to
characteristics of the set of functions $\{f_\theta : \theta \in
\Theta\}$, e.g., the `VC dimension' of that set~\citep[see][]{vapn82,vapn95}.

Another, historically earlier approach, is to apply bias-plus-variance
considerations to the $entire$ PL algorithm
${\underline{\hat{\theta}}}^\star$, rather than to each
$\Fhatbar(\theta)$ separately. This approach is applicable for
algorithms that do not use natural MCO, and even for non-parametric
supervised learning. As formulated for parameteric supervised
learning, this approach combines the formulas in
Eq.~\ref{eq:sup_learn_1} to write
\begin{eqnarray*}
F(\theta) &=& \int dx dy \; p(x) p(y \mid x) (z_\theta(x) - y)^2.
\end{eqnarray*}
This is then substituted into Eq.~\ref{eq:expectedLoss}, giving
\begin{eqnarray}
{\mean{[L]}} &=& \int d{\hat{\theta}}^\star  dx \,dy \; p(x)\, p(y \mid x)\,
p({\hat{\theta}}^\star) (z_{{{\hat{\theta}}}^\star}(x) - y)^2
\nonumber \\
&=& \int dx \; p(x) \left[ \int  d{\hat{\theta}}^\star dy \; p(x) p(y
\mid x)
p({\hat{\theta}}^\star) (z_{{{\hat{\theta}}}^\star}(x) - y)^2 \right].
\label{eq:sup_learn_bpv}
\end{eqnarray}
The term in square brackets is an $x$-parameterized
expected quadratic loss, which can be
decomposed into a bias, variance, etc., in the usual way. This formulation eliminates any direct concern for issues like
the distribution of extrema of multiple random variables, covariances
between $\Fhatbar(\theta)$ and $\Fhatbar(\theta')$ for different
values of $\theta$, and so on.

There are numerous other approaches for addressing the problems of
natural MCO that have been explored in PL. Particulary important among
these are Bayesian approaches, e.g.,
~\cite{buwe91,berg85,mack03}. Based on these approaches, as well as on intuition, many powerful techniques for addressing
data-overfitting have been explored in PL, including regularization,
cross-validation, stacking, bagging, etc. Essentially all
of these techniques can be applied to $any$ MCO problem, not just PL
problems. Since many of these techniques can be justified using
Eq.~\ref{eq:sup_learn_bpv}, they provide a way to exploit the
bias-variance trade-off in other domains besides PL.

\subsection*{PLMCO}
In this section, we illustrate how PL techniques that exploit the
bias-variance decomposition of Eq.~\ref{eq:sup_learn_bpv} can be used
to improve an MCO algorithm used in a domain outside of PL. This MCO
algorithm is a version of adaptive importance sampling, somewhat
similar to the CE method \citep{rukr04}, and is related to function
smoothing on continuous spaces. The PL techniques described are
applicable to any other MCO problem, and this particular one is chosen
just as an example.

\subsubsection*{MCO Problem Description}

Consider the problem of finding the $\theta$-parameterized distribution
$q_\theta$ that minimizes the associated expected value of a function
$G\colon \mathbb{R}^n \rightarrow \mathbb{R}$, i.e., find
\begin{equation*}
\arg \min_{\theta} \mean_{q_{\theta}}[G] .
\end{equation*}
We are interested in versions of this problem where we do not know the
functional form of $G$, but can obtain its value $G(x)$ at any $x \in
\mathcal{X}$. Similarly we cannot assume that $G$ is smooth, nor can
we evaluate its derivatives directly. This scenario arises in many
fields, including blackbox optimization \citep[see][]{wost06}, and
risk minimization
\citep[see][]{erno98}. 

We begin by expressing this minimization problem as an MCO problem. Write
\begin{eqnarray*}
\mean_{q_{\theta}}[G] &=& \int_{\mathcal{X}} dx\, q_{\theta}(x) G(x)
\end{eqnarray*}
Using MCO terminology, $f_{\theta}(x) = q_{\theta}(x) G(x)$
and $F(\theta) = \mean_{q_{\theta}}[G]$. To apply MCO, we must
define a vector-valued random variable $\Fhatbar$ with components
indexed by $\theta$, and then use a sample of $\Fhatbar$ to estimate
$\arg\min_\theta \mean_{q_{\theta}}[G]$. In particular, to apply naive
MCO to estimate $\arg\min_\theta \mean_{q_\theta}{(G)}$, we first i.i.d. sample a
density function $h(x)$. By evaluating the associated values of $G(x)$
we get a data set
\begin{eqnarray*}
{\mathcal{D}} &\equiv& ({\mathcal{D}}_{\cal{X}}, {\mathcal{D}}_G)
\nonumber \\
&=& (\{x^{(i)} : i = 1, \ldots, m\}, \{G(x^{(i)}) : i = 1, \ldots, m\}).
\end{eqnarray*}
The associated estimates of $F(\theta)$ for each $\theta$ are
\begin{eqnarray}
\Fhat(\theta) &\triangleq& \frac{1}{m}\sum_{i=1}^m
\frac{q_{\theta}(x^{(i)}) G(x^{(i)})}{h(x^{(i)})} .
\label{eq:fhat_1}
\end{eqnarray}
The associated naive MCO estimate of $\arg\min_\theta \mean_{q_{\theta}}[G]$ is
\begin{eqnarray*}
{\hat{\theta}}^{\star} &\equiv& \arg\min_{\theta}\hat{F}(\theta).
\end{eqnarray*}

Suppose $\Theta$ includes all possible density functions over
$x$'s. Then the $q_\theta$ minimizing our estimate is a delta function
about the $x^{(i)} \in {\mathcal{D}_{\mathcal{X}}}$ with the lowest associated value
of $G(x^{(i)})/h(x^{(i)})$. This is clearly a poor estimate in general; it
suffers from `data-overfitting'. Proceeding as in PL, one way to
address this data-overfitting is to use regularization. In particular,
we can use the entropic regularizer, given by the negative of the
Shannon entropy $S(q_{\theta})$. So we now want to find the minimizer of $\mean_{q_{\theta}}[G(x)] - T S(q_{\theta})$, where $T$ is the regularization parameter. Equivalently, we can minimize $\beta \mean_{q_{\theta}}[G(x)] - S(q_{\theta})$, where $\beta =1/T.$ This changes the definition of $\Fhat$ from the function
given in Eq.~\ref{eq:fhat_1} to
\begin{eqnarray*}
\Fhat(\theta) &\triangleq&
\frac{1}{m}\sum_{i=1}^m \frac{\beta \,q_{\theta}(x^{(i)}) G(x^{(i)})}{h(x^{(i)})} - S(q_{\theta}).
\end{eqnarray*}
Find the solution to this minimization problem is the focus of the PC
approach to blackbox optimization.

\subsubsection*{Solution Methodology}
Unfortunately, it can be difficult to find the $\theta$ globally
minimizing this new $\Fhat$ for an arbitrary $\cal{D}$. An alternative
is to find a close approximation to that optimal $\theta$. One way to
do this is as follows. First, we find the minimizer of 
\begin{equation}
\frac{1}{m}\sum_{i=1}^m \frac{\beta\, p(x^{(i)}) G(x^{(i)})}{h(x^{(i)})} - S(p)
\label{eq:minEGOverAllp}
\end{equation}
over the set of $all$ possible distributions $p(x)$ with domain ${\cal{X}}$. We then
find the $q_{\theta}$ that has minimal Kullback-Leibler (KL)
divergence from this $p$, evaluated over
${\mathcal{D}}_{\cal{X}}$. That serves as our approximation to
$\arg\min_\theta \Fhat(\theta)$, and therefore as our estimate of the
$\theta$ that minimizes $\mean_{q_\theta}{(G)}$.

The minimizer $p$ of Eq.~\ref{eq:minEGOverAllp} can be found in closed form; over ${\mathcal{D}}_{\cal{X}}$ it is the Boltzmann distribution $
p^{\beta}(x^{(i)}) \propto \exp(-\beta\,G(x^{(i)})).$ The KL divergence in ${\mathcal{D}}_{\cal{X}}$ from this Boltzmann
distribution to $q_{\theta}$ is
\begin{equation*}
F(\theta) = \textrm{KL}(p^{\beta}\|q_{\theta}) = \int_{\mathcal{X}}dx\,p^{\beta}(x)\log\left(\frac{p^{\beta}(x)}{q_{\theta}(x)}\right).
\end{equation*}
The minimizer of this KL divergence is given by
\begin{equation}
\theta^{\dagger} = \arg\min_{\theta} [-\sum_{i=1}^{m}\frac{\exp(-\beta G(x^{(i)}))}{h(x^{(i)})}\log(q_{\theta}(x^{(i)}))].
\label{eq:CEMinimization}
\end{equation}
$\theta^{\dagger}$ is an approximation to the estimate of the $\theta$
that minimizes $\mean_{q_\theta}{(G)}$ given by the regularized
version of naive MCO. Our incorporation of regularization here has the
same motivation as it does in PL: to reduce bias plus variance.

\subsubsection*{Log-concave Densities}
If $q_{\theta}$ is log-concave in its parameters $\theta$, then the minimization problem in Eq.~\ref{eq:CEMinimization} is a convex optimization problem, and the optimal parameters can be found closed-form. Denote the likelihood ratios by $s^{(i)} = \exp(-\beta G(x^{(i)}))/h(x^{(i)})$. Differentiating Eq.~\ref{eq:CEMinimization} with respect to the parameters $\mu$ and $\Sigma^{-1}$ and setting them to zero yields
\begin{eqnarray*}
\mu^{\star} &=& \displaystyle\frac{\sum_{\mathcal{D}} \displaystyle s^{(i)} x^{(i)}}{\sum_{\mathcal{D}} s^{(i)}}\\
\Sigma^{\star} &=& \frac{\sum_{\mathcal{D}} s^{(i)} (x^{(i)} - \mu^{\star})(x^{(i)} - \mu^{\star})^T}{\sum_{\mathcal{D}} s^{(i)}}
\end{eqnarray*}

\subsubsection*{Mixture Models}
The single Gaussian is a fairly restrictive class of models. Mixture models can significantly improve flexibility, but at the cost of convexity of the KL distance minimization problem. However, a plethora of techniques from supervized learning, in particular the Expectation Maximization (EM) algorithm, can be applied with minor modifications. 

Suppose $q_{\theta}$ is a mixture of $M$ Gaussians, that is, $\theta = (\mu, \Sigma, \phi)$ where $\phi$ is the mixing p.m.f, we can view the problem as one where a hidden variable $z$ decides which mixture component each sample is drawn from. We then have the optimization problem
\begin{equation*}
\textrm{minimize } -\sum_{\mathcal{D}} \displaystyle\frac{p(x^{(i)})}{h(x^{(i)})}\log\left(q_{\theta}(x^{(i)}, z^{(i)})\right).
\end{equation*}
Following the standard EM procedure, 
we get the algorithm described in Eq.~\ref{eq:EM}. Since this is a nonconvex problem, one typically runs the algorithm multiple times with random initializations of the parameters.
\begin{equation}
\label{eq:EM}
\begin{array}{lrcl}
\textrm{E-step: For each i, set} & Q_i(z^{(i)}) &=& p(z^{(i)}|x^{(i)}),\\
\hspace{0.5in}\textrm{that is,}      & w^{(i)}_j &=& q_{\mu, \Sigma, \phi}(z^{(i)} = j|x^{(i)}) ,\;\;\;j=1,\ldots,M.\\
\textrm{M-step: Set} & \mu_j &=& \frac{\sum_{\mathcal{D}} w^{(i)}_j  s^{(i)}\,x^{(i)}}{\sum_{\mathcal{D}} w^{(i)}_j \displaystyle s^{(i)}},\\
   &\Sigma_j &=& \frac{\sum_{\mathcal{D}} w^{(i)}_j \displaystyle s^{(i)}\,(x^{(i)} - \mu_j)(x^{(i)} - \mu_j)^T}{\sum_{\mathcal{D}} w^{(i)}_j  s^{(i)}},\\
&\phi_j &=& \frac{\sum_{\mathcal{D}} w^{(i)}_j s^{(i)}}{\sum_{\mathcal{D}} s^{(i)}},\\
\end{array}
\end{equation}

\subsubsection*{Test Problems}
To compare the performance of this algorithm with and without the use of PL techniques, we use a couple of very simple academic problems in two and four dimensions - the Rosenbrock function in two dimensions, given by
\begin{equation*}
G_R(x) = 100(x_2 - x_1^2)^2 + (1 - x_1)^2,
\end{equation*}
and the Woods function in four dimensions, given by
given by
\begin{eqnarray*}
G_{\mathrm{woods}}(x) &=& 100(x_2 - x_1)^2 + (1 - x_1)^2 + 90(x_4 - x_3^2)^2 + (1 - x_3)^2  \nonumber\\
     &&+ 10.1[(1 - x_2)^2 + (1 - x_4)^2] + 19.8(1-x_2)(1 - x_4).
\end{eqnarray*}
For the Rosenbrock, the optimum value of 0 is achieved at $x=(1,1)$, and for the Woods problem, the optimum value of 0 is achieved at $x = (1,1,1,1)$.

\subsubsection*{Application of PL Techniques}

As mentioned above, there are many PL techniques beyond regularization
that are designed to optimize the trade-off between bias and variance.
So having cast the solution of $\arg\min_{q_\theta} \mean{(G)}$ as an
MCO problem, we can apply those other PL techniques instead of (or in
addition to) entropic regularization. This should improve the
performance of our MCO algorithm, for the exact same reason that using
those techniques to trade off bias and variance improves performance
in PL. We briefly mention some of those alternative techniques here.

\begin{figure}
\centering
\begin{tabular}{cc}
\includegraphics[width=2.25in]{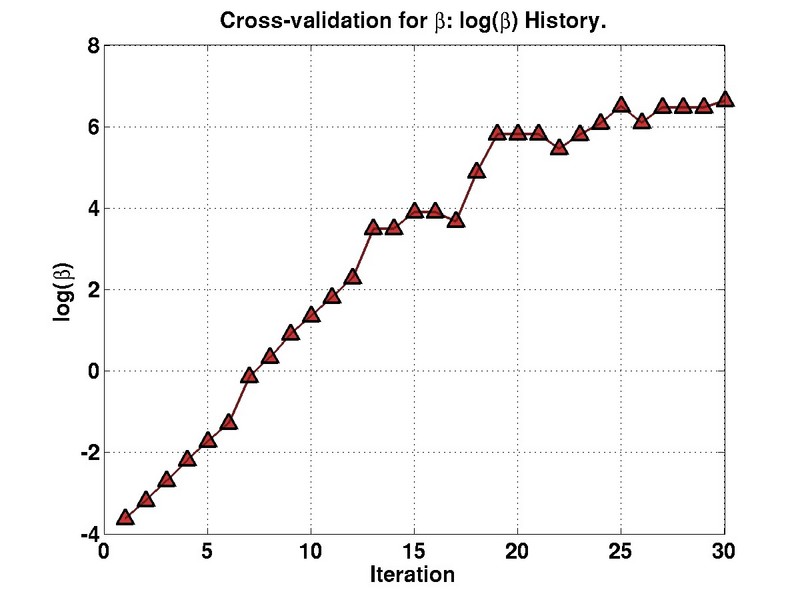} &
\includegraphics[width=2.25in]{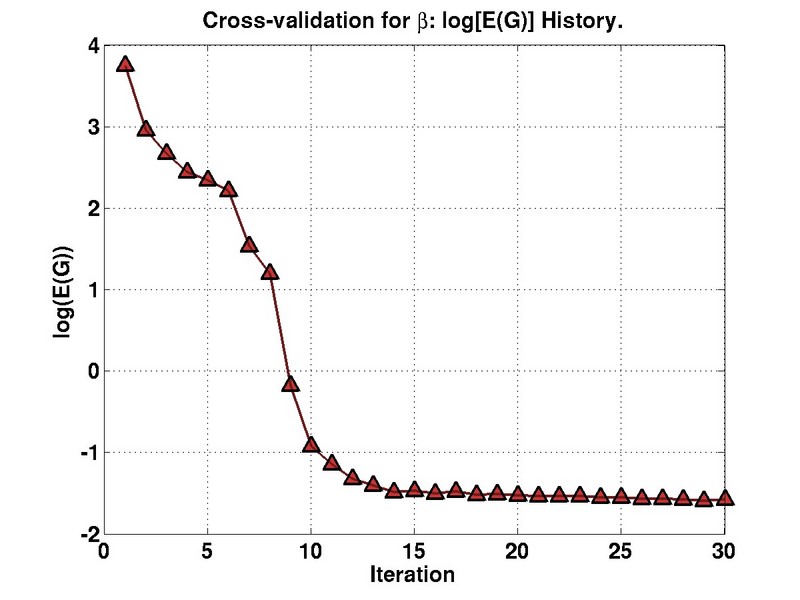}\\
a.&b.\\
\includegraphics[width=2.25in]{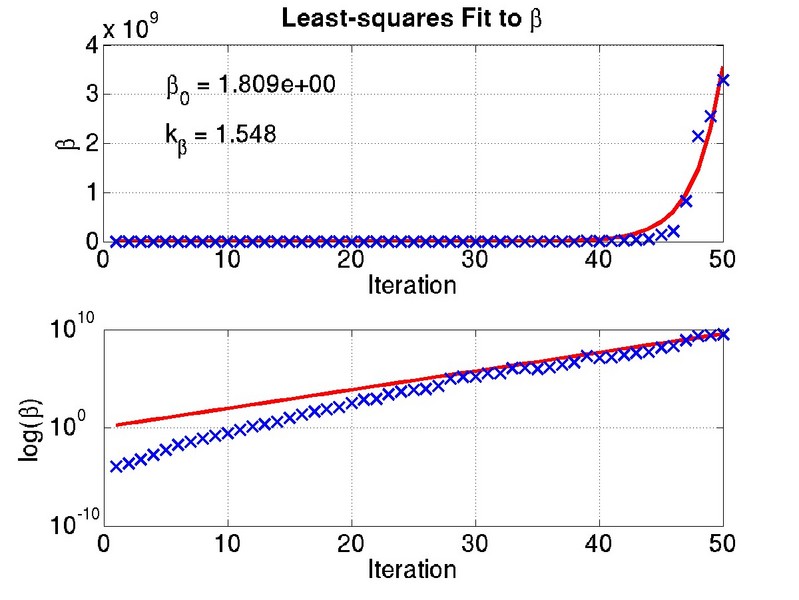} &
\includegraphics[width=2.25in]{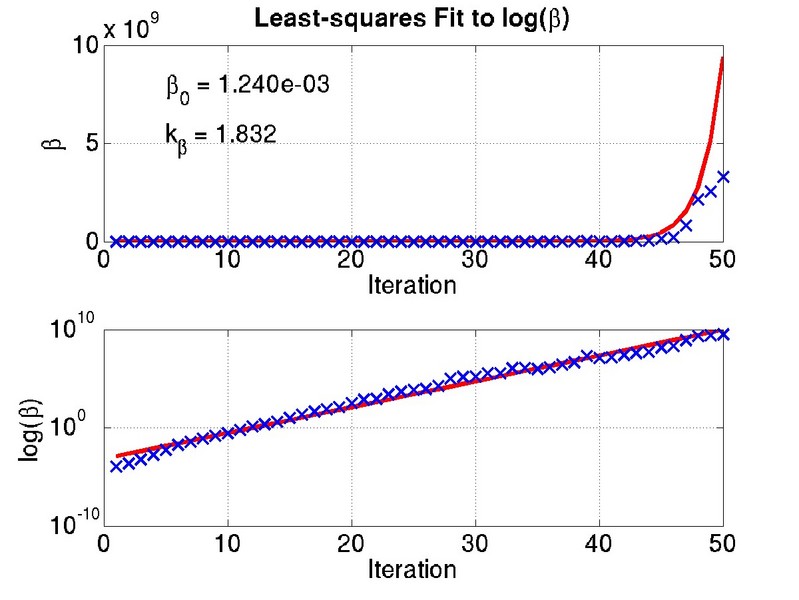}\\
c.&d.\\
\includegraphics[width=2.25in]{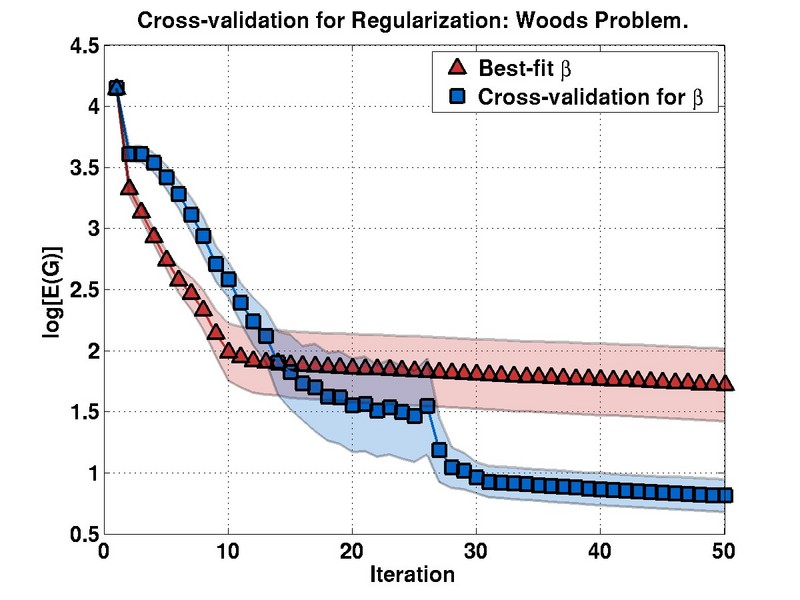} &
\includegraphics[width=2.25in]{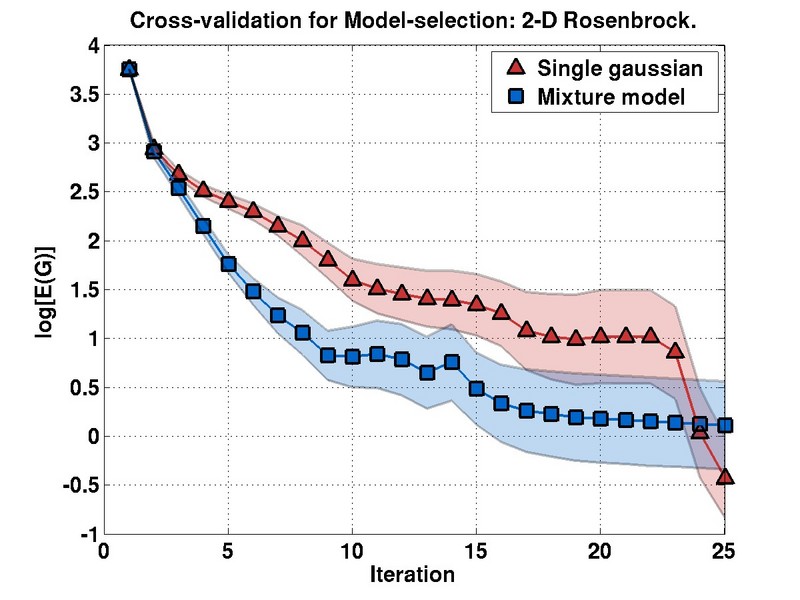}\\
e.&f.\\
\includegraphics[width=2.25in]{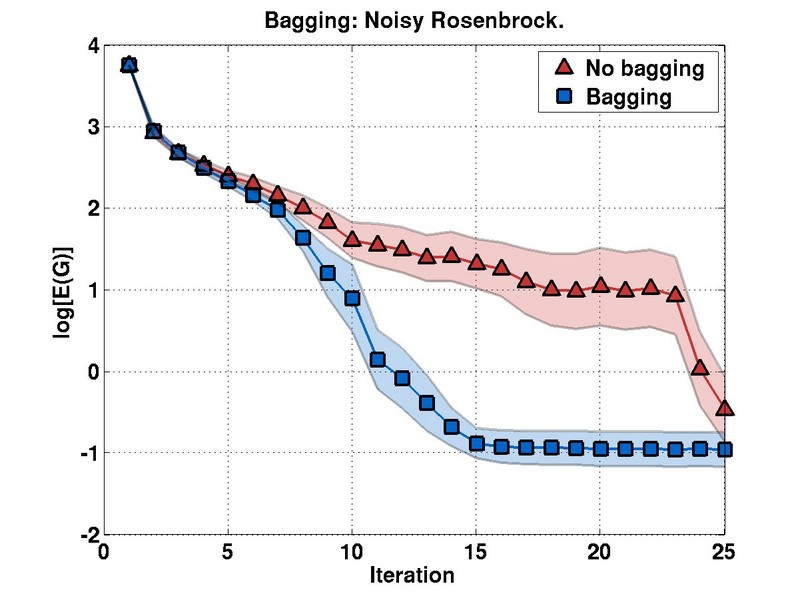} &
\includegraphics[width=2.25in]{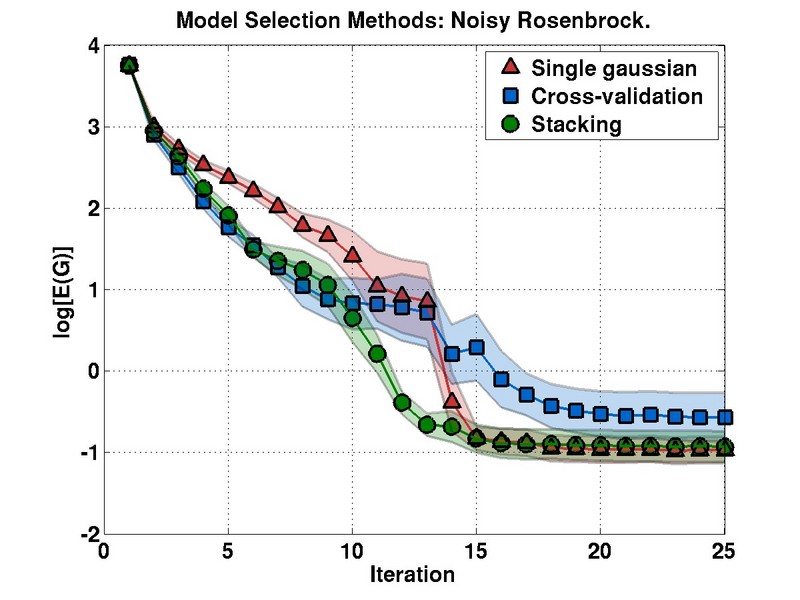} \\
g.&h.
\end{tabular}
\caption{Various PL Techniques Improve MCO performance.}
\label{fig:resultsPLMCO}
\end{figure}

The overall MCO algorithm is broadly described in Alg.~\ref{alg:gaussMixOverview}. For the Woods problem, 20 samples of $x$ are drawn from the updated $q_{\theta}$ at each iteration, and for the Rosenbrock, 10 samples. For comparing various methods and plotting purposes, 1000 samples of $G(x)$ are drawn to evaluate $\mean_{q_{\theta}}[G(x)]$. Note: in an actual optimization, we will not be drawing these test samples! All the performance results in Fig.~\ref{fig:resultsPLMCO} are based on 50 runs of the PC algorithm, randomly initialized each time. The sample mean performance across these runs is plotted along with 95\% confidence intervals for this sample mean (shaded regions).
\begin{algorithm}[h]
\caption{Overview of $pq$ minimization using Gaussian mixtures}
\label{alg:gaussMixOverview}
\begin{algorithmic}[1]
\STATE{Draw uniform random samples on $ X $}
\STATE{Initialize regularization parameter $\beta$}
\STATE{Compute $G(x)$ values for those samples}
\REPEAT
\STATE{Find a mixture distribution $q_{\theta}$ to minimize sampled $pq$ KL distance}
\STATE{Sample from $q_{\theta}$}
\STATE{Compute $G(x)$ for those samples}
\STATE{Update $\beta$}
\UNTIL{Termination}
\STATE{Sample final $q_{\theta}$ to get solution(s).}
\end{algorithmic}
\end{algorithm}

\noindent\textbf{Cross-validation for Regularization:} We note that we are using regularization to reduce variance, but that regularization introduces bias. As is done in PL, we use standard $k$-fold cross-validation to tradeoff this bias and variance. We do this by partitioning the data into $k$ disjoint sets. The held-out data for the $i^{\mathrm{th}}$ fold is just the $i^{\mathrm{th}}$ partition, and the held-in data is the union of all other partitions. First, we `train' the regularized algorithm on the held-in data $\mathcal{D}_t$ to get an optimal set of parameters $\theta^{\star}$, then `test' this $\theta^{\star}$ by considering unregularized performance on the held-out data $\mathcal{D}_v$. In our context, `training' refers to finding  optimal parameters by KL distance minimization using the held-in data, and `testing' refers to estimating $\mean_{q_{\theta}}[G(x)]$ on the held-out data using the following formula~\citep{roca04}.
\begin{equation*}
\widehat{g}(\theta) = \displaystyle\frac{\displaystyle\sum_{\mathcal{D}_v} \frac{q_{\theta}(x^{(i)}) G(x^{(i)})}{h(x^{(i)})}}{\displaystyle\sum_{\mathcal{D}_v} \frac{q_{\theta}(x^{(i)})}{h(x^{(i)})}}.
\end{equation*}
We do this for several values of the regularization parameter $\beta$ in the interval $k_1 \beta < \beta < k_2 \beta$, and choose the one that yield the best held-out performance, averaged over all folds. For our experiments, $k_1 = 0.5, k_2 = 3$, and we use 5 equally-spaced values in this interval. Having found the best regularization parameter in this range, we then use \emph{all} the data to minimize KL distance using this optimal value of $\beta$. Note that all cross-validation is done \emph{without} any additional evaluations of $G(x)$. Cross-validation for $\beta$ in PC is similar to optimizing the annealing schedule in simulated annealing. This `auto-annealing' is seen in Fig.~\ref{fig:resultsPLMCO}.a, which shows the variation of $\beta$ with iterations of the Rosenbrock problem. It can be seen that $\beta$ value sometimes decreases from one iteration to the next. This can never happen in any kind of `geometric annealing schedule', $\beta \leftarrow k_{\beta} \beta,\; k_{\beta}> 1$, of the sort that is often used in most algorithms in the literature. In fact, we ran 50 trials of this algorithm on the Rosenbrock and then computed a best-fit geometric variation for $\beta$, that is, a nonlinear least squares fit to variation of $\beta$, and a linear least squares fit to the variation of $\log(\beta)$. These are shown in Figs.~\ref{fig:resultsPLMCO}.c. and \ref{fig:resultsPLMCO}.d. As can be seen, neither is a very good fit. We then ran 50 trials of the algorithm with the fixed update rule obtained by best-fit to $\log(\beta)$, and found that the adaptive setting of $\beta$ using cross-validation performed an order of magnitude better, as shown in Fig.~\ref{fig:resultsPLMCO}.e.

\noindent\textbf{Cross-validation for Model Selection:}
Given a set $\Theta$ (sometimes called a model class) to choose $\theta$ from, we can find an optimal $\theta \in \Theta$. But how do we choose the set $\Theta$? In PL, this is done using cross-validation. We choose that set $\Theta$ such that $\arg\min_{\theta\in\Theta}\hat{F}(\theta)$ has the best held-out performance. As before, we use that model class $\Theta$ that yields the lowest estimate of $\mean_{q_{\theta}}[G(x)]$ on the held-out data. We demonstrate the use of this PL technique for minimizing the Rosenbrock problem, which has a long curved valley that is poorly approximated by a single Gaussian. We use cross-validation to choose between a Gaussian mixture with up to 4 components. The improvement in performance is shown in Fig.~\ref{fig:resultsPLMCO}.d. 

\noindent\textbf{Bagging:} In bagging \cite{brei96a}, we generate multiple data sets by resampling the given data set with replacement. These new data sets will, in general, contain replicates. We `train' the learning algorithm on each of these resampled data sets, and average the results. In our case, we average the $q_{\theta}$ got by our KL divergence minimization on each data set. PC works even on stochastic objective functions, and on the noisy Rosenbrock, we implemented PC with bagging by resampling 10 times, and obtained significant performance gains, as seen in Fig.~\ref{fig:resultsPLMCO}.g.

\noindent\textbf{Stacking:} In bagging, we combine estimates of the same learning algorithm on different data sets generated by resampling, whereas in stacking \cite{brei96, smwo99}, we combine estimates of different learning algorithms on the same data set. These combined estimated are often better than any of the single estimates. In our case, we combine the $q_{\theta}$ obtained from our KL divergence minimization algorithm using multiple models $\Theta$. Again, Fig.~\ref{fig:resultsPLMCO}.h shows that cross-validation for model selection performs better than a single model, and stacking performs slightly better than cross-validation.

\subsection*{Conclusions}
The conventional goal of reducing bias plus variance has interesting
applications in a variety of fields. In straightforward applications,
the bias-variance trade-offs can decrease the MSE of estimators, reduce
the generalization error of learning algorithms, and so on. In this
article, we described a novel application of bias-variance trade-offs:
we placed bias-variance trade-offs in the context of Monte Carlo
Optimization, and discussed the need for higher moments in the
trade-off, such as a bias-variance-covariance trade-off. We also showed
a way of applying just a bias-variance trade-off, as used in Parametric
Learning, to improve the performance of Monte Carlo Optimization
algorithms.  

\bibliography{dgorurBib}

\end{document}